\documentclass[twocolumn,pra,aps]{revtex4-1}
\usepackage{graphicx,hyperref}
\begin{document}
\title{ A classical optical approach to the `non-local
Pancharatnam-like phases' in
Hanbury-Brown-Twiss correlations}
\author{Arvind}
\email{arvind@iisermohali.ac.in}
\affiliation{Department of Physical Sciences,
Indian
Institute of Science Education \&
Research (IISER) Mohali, Sector 81 SAS Nagar,
Manauli PO 140306 Punjab India.}
\author{ S. Chaturvedi}
\email{subhash@iiserbhopal.ac.in}
\affiliation{Department of Physics, ,
Indian
Institute of Science Education \&
Research (IISER) Bhopal, Bhopal Bypass Road, Bhauri, Bhopal 462066 India}
\author{N. Mukunda}
\email{nmukunda@gmail.com}
\affiliation{
INSA C V Raman Research Professor,
Indian Academy of Sciences,
C V Raman Avenue, Sadashivanagar,
Bangalore 560080 India}
\begin{abstract} We examine a recent proposal to
show the presence of nonlocal Pancharatnam type
geometric phases in a quantum mechanical treatment
of intensity interferometry measurements upon
inclusion of polarizing elements in the setup.  It
is shown that a completely classical statistical
treatment of such effects is adequate for
practical purposes. Further we show that the phase
angles that appear in the correlations, while at
first sight appearing to resemble Pancharatnam
phases in their mathematical structure, cannot
actually be interpreted in that manner. We also
describe a simpler Mach-Zehnder type setup where
similar effects can be observed without use of the
paraxial approximation. 
\end{abstract}
\maketitle
\section{Introduction}
\label{intro}
The work of Hanbury-Brown and Twiss (HBT) about six
decades ago inaugurating the field of intensity
interferometry in radio astronomy as well as in
the visible region constituted a major conceptual
and experimental advance in the subject~\cite{HBT}.  Even
though initially there was some confusion
regarding interpretation, especially with regard
to the quantum mechanical meaning of HBT
correlations, it has since been recognized that it
can be satisfactorily understood in terms of the
statistical features of general states of
classical (optical) wave fields. Each such
statistical state is describable by a hierarchy of
correlation functions of various orders, and the
HBT intensity-intensity correlation function
stands one step beyond the more familiar Young
type amplitude-amplitude correlation function
(also called the two-point function describing
partial coherence) adequate for handling
interference and diffraction phenomena. As with
the Bell inequalities which characterize proposed
local realistic extensions of quantum mechanics, and
which can be violated by specific entangled
quantum states, in the HBT case too particular
quantum states of radiation may lead to
correlations beyond what classical theory can
explain. However this does not invalidate the fact
that as a concept the HBT correlations are
classically meaningful.

In contrast to the HBT effect, the concept of
geometric phases in quantum mechanics was uncovered
by Berry just over three decades ago~\cite{berry}. 
His analysis
was in the framework of adiabatic cyclic unitary
evolution
%page 3
of pure quantum states obeying the time-dependent
Schr\"{o}dinger equation - at the end of such
evolution the state vector (or wave function) in
Hilbert space acquires a new previously
unrecognized phase. Later rapid developments
greatly clarified the situation - the geometric
phase is (in the language of quantum mechanics) a
ray space quantity; it can be defined even in
nonadiabatic and noncyclic
evolutions~\cite{anandan,sambhandari,wilczek}; and it is
meaningful in purely classical wave optical
situations, so it is not specifically quantum
mechanical in origin. Indeed it  was soon
realised that a phase found by Pancharatnam in
1956 in classical polarization optics was an early
precursor of the geometric phase in a nonadiabatic
cyclic situation, with the Poincar\'{e} sphere of
polarization optics playing the role of ray space
in quantum
mechanics~\cite{pancharatnam,rajaramramaseshan,berrypanch}.

Subsequent work on the kinematic approach to the
geometric phase has shown that the basic
ingredient is the use of a complex Hilbert space
to describe (pure) states of a physical system,
whether in quantum mechanics or in classical wave
optics, and the associated ray
space~\cite{kinematic1,kinematic2}. It has also
brought out the relevance of the Bargmann
invariants for geometric phase
theory~\cite{bargmann}. While some
attempts have been made to define geometric phases
for mixed state evolution~\cite{Uhl,Sjo,Cha,And}, the emphasis at the
basic level has been on pure states.

Against this background, some very interesting
recent work has attempted to bring together these
two independent developments in  an unexpected
manner~\cite{sam_nl}. It has been shown that in a carefully
prepared experimental setup involving polarizing
gadgets the expression for HBT correlations
contains just the kind of phase angle - a solid
angle on a two-sphere - involved in
%Page 4
Pancharatnam's work. This has been described as a
nonlocal form of the Pancharatnam phase, and it
has been subjected to an experimental test as
well~\cite{Martin,Sat}. In particular the theoretical analysis uses
the photon description of light, involving
quantized radiation field operators; and the
nonlocal effect has been characterized as a
genuinely two-photon property not visible at the
single photon level.

The aims of the present work are two-fold :  the
first is to show that a purely classical
statistical treatment of radiation is adequate to
obtain the result of~\cite{sam_nl}, without having
to use the photon picture based on the quantum
theory of radiation; the second is to
significantly simplify the experimental set up
while retaining the appearance of the Pancharatnam
solid angle in the expression for HBT correlations,
and so to understand better whether it is indeed a
nonlocal form of the Pancharatnam phase which in
any event is a particular instance of the
geometric phase.

The contents of the paper are arranged as follows:
Section~\ref{hbt} describes the original HBT
scheme of~\cite{sam_nl} from a purely classical
point of view. In Section~\ref{mz} we give a
scheme based on a Mach-Zender interferometer,
where similar results are obtained without the
paraxial approximation. Section~\ref{conc} offers
some concluding remarks and the Appendix
describes the mathematical conventions used in
this paper.
%%%%%%%%%%%%%%%%%
%page 5
\section{Classical treatment of HBT correlations
with polarizers}
\label{hbt}
In this Section we present a completely classical
treatment of the HBT correlations involving the
Pancharatnam solid angle, comparing
with~\cite{sam_nl}
at relevant points.

\begin{figure}
\includegraphics[scale=1]{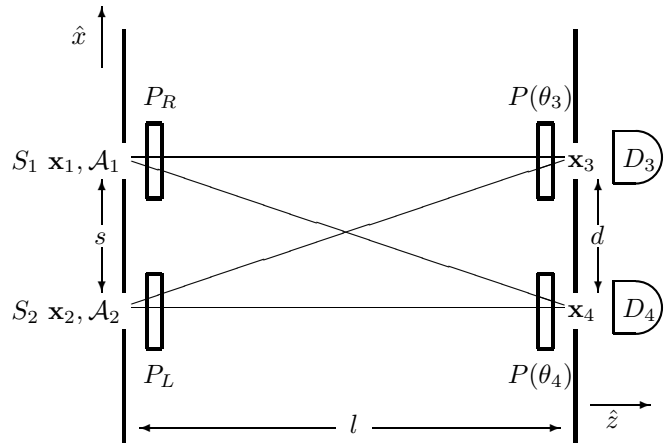}
\caption{Scheme for HBT correlations with
polarizers.}
\label{hbt_setup}
\end{figure}

We assume two localized sources $S_1, S_2$ of
quasimonochromatic radiation of mean frequency
$\omega_0$, in independent statistical states (more
fully specified below). As in
Fig.~\ref{hbt_setup}, they are located a distance
$s$ apart, at points $\bf{x}_1,
\bf{x}_2$ along the x-axis of a spatial
coordinate system. Light from each source reaches
two detectors $D_3$ and $D_4$ at positions
$\bf{x}_3, \bf{x}_4$ a distance $d$
apart, and at a distance $l$ from the sources in
the overall direction of the positive z-axis, with
$l >> s,d$. Therefore the propagation vectors of
light waves from $S_1, S_2$ to $D_3, D_4$ may all
be treated as practically parallel and along the
positive z-axis, i.e. we are in the paraxial
regime. Polarizers $P_R, P_L$ placed immediately
after $S_1, S_2$ select right and left circular
polarizations respectively. Just before reaching
detector $D_3$, the superposed fields from $S_1$
and $S_2$ pass through a linear polarizer
$P(\theta_3)$ at an angle $\theta_3$ in the
transverse x-y plane; similarly another linear
polarizer $P(\theta_4)$ at angle $\theta_4$ is
placed just before $D_4$.  Within the limits of
the paraxial approximation all relevant electric
field vectors can be taken to be two-component
objects in the common transverse plane, with only
$x$ and $y$ components. 

In the absence of polarizers $P_R, P_L,
P(\theta_3), P(\theta_4)$, using the Kirchoff and
paraxial approximations the positive frequency
analytic signal electric field vectors reaching
$D_3$ from $S_a, a=1,2,$ are given by
\begin{eqnarray}
&&\underline{E}^{(a)}({\bf x}_3,t) \approx
u_{a3}
\underline{E}^{(a)}\left({\bf x}_a,t-
\frac{\mathcal{R}_{a3}}{c}\right),
\nonumber \\
&&u_{a3} = \frac{-i k_0 l }{2
\pi}\frac{\mathcal{A}_a}{\mathcal{R}^2_{a3} },
\mathcal{R}_{a3}= \vert
{\bf x}_a-{\bf x}_3\vert,
k_0=\frac{\omega_0}{c},\, a=1,2\nonumber \\
\label{field_amplitude}
\end{eqnarray}
Here $\mathcal{A}_a$ is the effective area of the
source $S_a$, $u_{a3}$ is a dimensionless
geometrical propagation factor from $S_a$ to $D_3$, 
and
$
\underline{E}^{(a)}\left({\bf x}_a,t-
\mathcal{R}_{a3}/c\right) 
$ is the electric field vector at $S_a$ at the
retarded time. We treat this field as  effectively 
constant over $\mathcal{A}_a$.  With the polarizers
in place as in Fig.~\ref{hbt_setup}, the total
field reaching $D_3$ is 
\begin{eqnarray}
\underline{E}({\bf x}_3,t)\approx&&
P(\theta_3)\left( 
u_{13} P_R
\underline{E}^{(1)}\left({\bf x}_1,t-
\mathcal{R}_{13}/c\right)
\right.
\nonumber \\
&&
\quad\quad\,
\left.
+
u_{23} P_L
\underline{E}^{(2)}\left({\bf x}_2,t-
\mathcal{R}_{23}/c\right)
\right).
\label{total_elec_field}
\end{eqnarray}
For $\underline{E}({\bf x}_4,t)$ at $D_4$ we
have a similar expression by replacing
${\bf x}_3 \rightarrow {\bf x}_4$,
$u_{a3} \rightarrow u_{a4}$, 
$\mathcal{R}_{a3} \rightarrow \mathcal{R}_{a4}$,
$P(\theta_3) \rightarrow P(\theta_4)$.

The intensities at $D_3$, $D_4$ are proportional
to $\underline{E}^{\star}({\bf x}_3,t)\cdot
\underline{E}({\bf x}_3,t)$ and
%page 7
$
\underline{E}^{\star}({\bf x}_4,t)\cdot
\underline{E}({\bf x}_4,t)$
respectively.
Since we are concerned with a completely classical discussion, 
it is appropriate to treat the fields $
\underline{E}^{(a)}({\bf x}_a,t)\equiv
\underline{E}^{(a)}$
as belonging to corresponding statistical ensembles, described 
by appropriate classical probability distributions. For our 
purposes these are represented by correlation functions of 
various orders~\cite{Klau,Wolf}.  (The time
arguments will be seen to be irrelevant, and as $l
>> s,d$ the differences in retarded times will be
neglected). Denoting ensemble averages by $\langle 
\cdots \rangle$, the HBT correlation of
intensities at $D_3$ and $D_4$ is proportional to 
\begin{eqnarray}
&&\Gamma^{(2,2)}=
\langle 
\underline{E}^{\star}{({\bf x}_3,t)}\cdot
\underline{E}{({\bf x}_3,t)}
\underline{E}^{\star}{({\bf x}_4,t)}\cdot
\underline{E}{({\bf x}_4,t)}
\rangle =\nonumber \\
&&\langle( 
u^{\star}_{13}\underline{E}^{(1)^{\dagger}} P_R+
u^{\star}_{23}\underline{E}^{(2)^{\dagger}} P_L)
P(\theta_3)
(u_{13} P_R\underline{E}^{(1)} +
u_{23} P_L \underline{E}^{(2)}).
\nonumber \\
&&(u^{\star}_{14}\underline{E}^{(1)^{\dagger}} P_R+
u^{\star}_{24}\underline{E}^{(2)^{\dagger}} P_L)
P(\theta_4)
(u_{14} P_R \underline{E}^{(1)}+
u_{24} P_L \underline{E}^{(2)})
\rangle  \nonumber \\
\label{gamma22}
\end{eqnarray}
This is a sum of sixteen terms, each a product of
four factors, which can be labelled in sequence as
$1111, 1112, \cdots , 2221, 2222$ according to
which term from each of the four bracketed
expressions is included in the product. Thus for
example
\begin{eqnarray}
1111 &=& 
\vert u_{13} \vert^2 
\vert u_{14} \vert^2. 
\nonumber \\
&&
\langle 
\underline{E}^{(1)^{\dagger}} 
P_R P(\theta_3) P_R 
\underline{E}^{(1)} 
\underline{E}^{(1)^{\dagger}} 
P_R P(\theta_4) P_R 
\underline{E}^{(1)} 
\rangle ,\nonumber \\
1112 &=& 
\vert u_{13} \vert^2 
u_{14}^{\star} u_{24}.
\nonumber \\
&&
\langle 
\underline{E}^{(1)^{\dagger}} 
P_R P(\theta_3) P_R 
\underline{E}^{(1)} 
\underline{E}^{(1)^{\dagger}} 
P_R P(\theta_4) P_L 
\underline{E}^{(2)} 
\rangle.
\label{gamma22_terms}\nonumber \\
\end{eqnarray}
We now specify the statistics of the source fields
$\underline{E}^{(a)}$. We make the physically
plausible assumption that $\underline{E}^{(1)}$
and 
$\underline{E}^{(2)}$ belong to two independent
time stationary centered random phase unpolarized
Gaussian ensembles. Recalling that
$\underline{E}^{(1)}$,~$\underline{E}^{(2)}$ are
two-component complex column vectors with $x$ and
$y$ components labelled by $\alpha, \beta = 1,2$,
we have the basic ensemble averages
%page 8
\begin{eqnarray}
&&
\langle \underline{E}^{(a)}_{\alpha} \rangle
=\langle \underline{E}^{(a)\star}_{\alpha} \rangle
=0, \nonumber \\
&&\langle \underline{E}^{(a)}_{\alpha} 
\underline{E}^{(a)}_{\beta} 
\rangle =
\langle \underline{E}^{(a)\star}_{\alpha} 
\underline{E}^{(a)\star}_{\beta} 
\rangle=0,\,\, a=1,2;
\nonumber \\
&&\langle \underline{E}^{(1)\star}_{\alpha} 
\underline{E}^{(1)}_{\beta} 
\rangle = \kappa\delta_{\alpha \beta},\,\,
\langle \underline{E}^{(2)^\star}_{\alpha} 
\underline{E}^{(2)}_{\beta} 
\rangle = \kappa^{\prime}\delta_{\alpha \beta};
\label{ensemble_properties1} 
\end{eqnarray}
and the derived averages
\begin{eqnarray}
\langle \underline{E}^{(1)^\star}_{\alpha^{\prime}} 
\underline{E}^{(1)^\star}_{\beta^{\prime}} 
\underline{E}^{(1)}_{\alpha} 
\underline{E}^{(1)}_{\beta} 
\rangle 
= \kappa^{2} (
\delta_{\alpha^{\prime} \alpha}
\delta_{\beta^{\prime} \beta}
+\delta_{\alpha^{\prime} \beta}
\delta_{\beta^{\prime} \alpha}
),
\nonumber \\ 
\langle \underline{E}^{(2)^\star}_{\alpha^{\prime}} 
\underline{E}^{(2)^\star}_{\beta^{\prime}} 
\underline{E}^{(2)}_{\alpha} 
\underline{E}^{(2)}_{\beta} 
\rangle 
= \kappa^{\prime 2} (
\delta_{\alpha^{\prime} \alpha}
\delta_{\beta^{\prime} \beta}
+\delta_{\alpha^{\prime} \beta}
\delta_{\beta^{\prime} \alpha}
).\nonumber \\
\label{ensemble_properties2} 
\end{eqnarray}
Here $\kappa$, $\kappa^{\prime}$ are in general different
real positive parameters. These expressions can be
easily reproduced by a suitable centered Gaussian
probability distribution for the four complex
amplitudes $\{ \underline{E}^{(a)}_\alpha \}$ treating all
polarizations uniformly. 

With these assumptions on the sources $S_1$, $S_2$
there are only six nonvanishing terms in
$\Gamma^{(2,2)}$ corresponding to the products 
$1111$, $1122$, $1221$, $2112$, $2211$ and $2222$.
These can be easily computed using the basic
properties of the $2\times 2 $ polarization
matrices $P_R$,$P_L$, $P(\theta_3)$, $P(\theta_4)$
given in the Appendix. We find:
\begin{eqnarray}
&&1111= 
\vert u_{13} \vert^2 
\vert u_{14} \vert^2 \frac{\kappa^{{2}}}{2};
\nonumber \\
&&1122= 
\vert u_{13} \vert^2
\vert u_{24} \vert^2 \frac{\kappa \kappa^{\prime}}{4};
\nonumber \\
&&1221=
u_{13}^{\star} u_{23}^{\phantom \star}
u_{24}^{\star} u_{14}^{\phantom \star}
\kappa \kappa^{\prime} {\rm Tr}(P_R
P(\theta_3)P_L(\theta_4)); \nonumber \\
&&2112=(1221)^{\star};
\nonumber \\
&&2211= 
\vert u_{23} \vert^2
\vert u_{14} \vert^2 \frac{\kappa
\kappa^{\prime}}{4};
\nonumber \\
&&2222= 
\vert u_{23} \vert^2 
\vert u_{24} \vert^2 \frac{\kappa^{\prime^{2}}}{2};
\nonumber \\
&&\Gamma^{(2,2)} = 
\frac{1}{2} (\vert u_{13} u_{14} \vert^2 \kappa^2
+\vert u_{23} u_{24} \vert^2 \kappa^{\prime^2})\nonumber \\
&& \quad \quad + \frac{1}{4} 
\left(\vert u_{13} u_{24} \vert^2 +
\vert u_{14} u_{23} \vert^2\right)\kappa \kappa^{\prime} 
\nonumber \\
&&\quad \quad
+ 2 \kappa \kappa^{\prime} {\rm Re}\left[u_{13}^{\star}
u_{23}^{\phantom \star} 
u_{14}^{\phantom \star} u_{24}^{\star} {\rm Tr}\left(P_R
P(\theta_3)P_LP(\theta_4) \right)
\right]. \nonumber \\
\label{gamma_in_terms}
\end{eqnarray}
 The last term in $\Gamma^{(2,2)}$ is reminiscent
of the Pancharatnam phase in the geometric phase
context. As shown in the Appendix, the trace term
has 
a phase related to the solid angle of a `lune' on
a two-sphere enclosed by the two meridians at polar
angles $2\theta_3$ and $2\theta_4$,
\begin{equation}
{\rm Tr}(P_R P(\theta_3) P_L P(\theta_4))
=\frac{1}{4} e^{-2i (\theta_3-\theta_4)},
\end{equation} 
so after including the propagation factors $u$ in
Eqn.(\ref{field_amplitude}) the last term in
$\Gamma^{(2,2)}$ becomes 
\begin{equation}
\frac{\kappa \kappa^{\prime}}{2} \left(\frac{k_0 l}{2\pi}
\right)^4 \left(\mathcal{A}_1
\mathcal{A}_2/
\mathcal{R}_{13} \mathcal{R}_{23} 
\mathcal{R}_{14} \mathcal{R}_{24}
\right)^2
\cos{2(\theta_3-\theta_4)}.
\label{geom_contrib_gamma}
\end{equation}

This is the contribution made by the presence of
the polarizers to the HBT correlations in the setup
of Fig.~\ref{hbt_setup}.

It is clear that the result~(\ref{geom_contrib_gamma})
depends strongly on the statistical
properties~(\ref{ensemble_properties1},\ref{ensemble_properties2})
assumed for the sources $S_1$ and $S_2$. Even a
slight change in them can lead to the final
result for $\Gamma^{(2,2)}$ not being expressible 
in terms of any recognizable solid angle on a
two-sphere at all. Thus the fact that the Pancharatnam
solid angle has entered the
result~(\ref{geom_contrib_gamma}) seems to be not
generic or robust.

We point out that the quantum mechanical treatment
in~\cite{sam_nl} involves defining photon annihilation
and creation operators $a_3, a_3^{\dagger}, a_4,
a_4^{\dagger}$ corresponding to the modes `at the
detectors' $D_3$, $D_4$ in terms of operators 
 $a_1, a_1^{\dagger}, a_2,
a_2^{\dagger}$ for modes `at the sources' $S_1$,
$_2$. The proposed relationships are exactly
parallel to Eqn.~(\ref{total_elec_field}) above
for classical
analytic signals and read:
\begin{eqnarray}
a_3&=& P(\theta_3) (u_{13} P_R a_1+ u_{23}P_L a_2 ),
\nonumber \\
a_4&=& P(\theta_4) (u_{14} P_R a_1+ u_{24}P_L a_2 ),
\end{eqnarray}
and their adjoints for $a_3^{\dagger}$,
$a_4^{\dagger}$. Here $a_3$ is a column vector of
two annihilation operators for the two photon
polarization states `at $D_3$', $a_3^{\dagger}$ is a 
row vector, etc.

However, using the classical wave propagation
formulae for photon operators in this way cannot
be expected to preserve the bosonic commutation
relations, and the meaning of $\langle
a_3^{\dagger}a_3 \rangle,  \langle a_4^{\dagger}
a_4\rangle$ in terms of photon numbers. A proper
quantum mechanical treatment needs to address
these points in a satisfactory manner.

Nevertheless, the fact that the use of polarizers in
HBT correlation measurements leads to nontrivial
consequences, while not surprising, is interesting
in itself. The proposed interpretation in the
language of Pancharatnam( and geometric) phases is
examined more closely in the next Section.

\section{An alternative experimental scheme}
\label{mz}
The paraxial approximation used in the previous
Section suggests that a common Poincar\'{e} sphere
could be used to describe the polarization states
of transverse electric fields propagating in any
of the four practically parallel directions $S_1$,
$S_2$ to $D_3$, $D_4$. However it should also be
mentioned that the analysis did not require
following the evolution of the polarization state
of any electric field two-vector along any closed
circuit on a Poincar\'{e} sphere, which is usually a
prerequisite to identify a Pancharatnam phase.

To understand the situation better we now consider
a simpler scheme involving only two
\underline{mutually orthogonal} propagation
directions, thus giving up the paraxial property
altogether.  This is similar to  but considerably
simpler than the one discussed in~\cite{Sat}.
We should point out that in general,
when there are plane waves simultaneously
propagating in several different directions, one
should in principle use a separate Poincar\'{e} sphere
attached to each propagation direction to follow the
polarization state of the field vector propagating
in that direction. Furthermore, for each
propagation direction there is an independent
$SO(2)$ freedom in the choice of transverse axes,
and even a $U(2)$ freedom in the choice of
orthonormal polarization states, in setting up
these individual Poincar\'{e} spheres. It is only when
there are physically significant simplifying
features, such as the paraxial condition for
instance, that many different Poincar\'{e} spheres may
be identified with one another.

\begin{figure}
\includegraphics[scale=1]{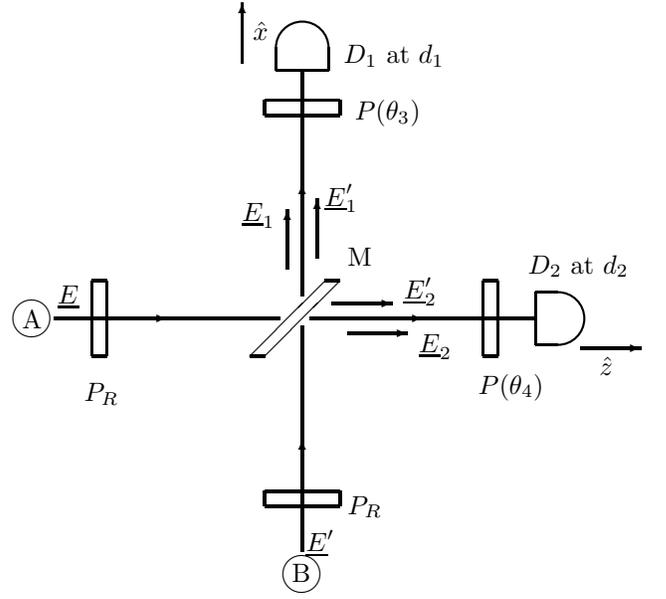}
\caption{A setup for HBT correlations in the
nonparaxial condition.}
\label{mz_setup}
\end{figure}

We consider a setup with two sources or input
ports $A,B$ producing plane electromagnetic waves
of common frequency $\omega_0$, propagating along
the positive $z$ and $x$ axes respectively, as in
Fig.~\ref{mz_setup} (the $y$ axis then points out of the
page). These waves pass through circular
polarizers $P_R$ located close to the sources. The
field from $A$ is described by a two-component
complex column vector $\underline{E}$ made up of
its $x$ and $y$ components, while
$\underline{E}^{\prime}$ from $B$ is similarly
described by its $y$ and $z$ components:
\begin{equation}
\underline{E} = \left( \begin{array}{c}
E_x \\
E_y \\
\end{array}
\right),
\quad 
\underline{E}^{\prime} = \left( \begin{array}{c}
E^{\prime}_{y} \\
E^{\prime}_{z}
\end{array}\right)
\label{column_fields}
\end{equation}
This is a natural convention for the present 
non-paraxial situation.
The plane wave factors are $e^{i k_0 z}$ for
$\underline{E}$ and $e^{i k_0 x}$ for
$\underline{E}^{\prime}$, but in the sequel these
can be omitted.

After the polarizers $P_R$ the two plane waves
arrive at a half-silvered plane mirror $M$ placed
at an angle of $45^{0}$ bisecting the $z$ and $x$
axes. Each plane wave is partially transmitted and
partially reflected by $M$, the intensities
divided equally. Thus $\underline{E}$ gives rise
to reflected $\underline{E}_1$ and transmitted
$\underline{E}_2$, while $\underline{E}^{\prime}$
gives rise to transmitted
$\underline{E}_1^{\prime}$ and reflected
$\underline{E}_2^{\prime}$. Both
$\underline{E}_1^{\prime}$ and $\underline{E}_1$
are described like $\underline{E}^{\prime}$
in~(\ref{column_fields}), $\underline{E}_2$ and
$\underline{E}_2^{\prime}$ like $\underline{E}$.
At the mirror, the condition that there be no
tangential field (and the stated reduction in intensity) determines $\underline{E}_1$ in
terms of $\underline{E}$, and
$\underline{E}_2^{\prime}$ in terms of
$\underline{E}^{\prime}$. Keeping in mind
the conventions~(\ref{column_fields}), it is seen that mirror
reflection is expressed by the matrix $\tau_2$ (Pauli
matrix $\sigma_1$ as explained in
Eqn.~(\ref{pol_pauli})) . Since we are dealing with
circular polarizations, upon reflection RCP and
LCP get interchanged, while upon transmission they
each remain unchanged. All in all, the two sets of
reflected and transmitted waves are:
\begin{eqnarray} 
\mbox{Reflected~~~} 
\underline{E}_1
= - \frac{1}{\sqrt{2}} \tau_2 P_R \underline{E},
\quad \underline{E}_2^{\prime} =
-\frac{1}{\sqrt{2}} \tau_2 P_R
\underline{E}^{\prime} ;\nonumber \\
\mbox{Transmitted~~} 
\underline{E}_2 =
\frac{1}{\sqrt{2}} P_R
\underline{E}.\quad
\underline{E}_1^{\prime} = \frac{1}{\sqrt{2}} P_R
\underline{E}^{\prime}. 
%\nonumber \\ 
\label{pol_tau1}
\end{eqnarray}
The projection matrix $P_R$ of
Eq.~(\ref{polarizer_rep}) can be used
consistently for both $\underline{E}$ and
$\underline{E}^{\prime}$, using the conventions(\ref{column_fields}).

After superposition,
$\underline{E}_1+\underline{E}_1^{\prime}$ passes
through a linear polarizer $P(\theta_3)$ (in the
$y-z$ plane), while
$\underline{E}_2+\underline{E}_2^{\prime}$ passes
through $P(\theta_4)$ (in the $x-y$ plane). Each
of these superpositions consists of one RCP wave
and one LCP wave. They are then received at
detectors $D_1$, $D_2$ respectively, after having
traversed distances $d_1$ and $d_2$ from the
mirror $M$. The total fields reaching the
detectors are: \begin{eqnarray} \underline{E}(D_1)
&=& \frac{1}{\sqrt{2}} P(\theta_3) (- \tau_2 P_R
\underline{E} + P_R \underline{E}^{\prime}),
\nonumber \\ \underline{E}(D_2) &=&
\frac{1}{\sqrt{2}} P(\theta_4) (P_R \underline{E}
- \tau_2 P_R \underline{E}^{\prime}) .\nonumber
\\ 
\label{pol_tau2}
\end{eqnarray} Therefore the HBT correlation of
the two intensities is \begin{eqnarray}
\Gamma^{(2,2)} &=& \langle
\underline{E}(D_1)^{\dagger} \underline{E}(D_1)
\underline{E}(D_2)^{\dagger} \underline{E}(D_2)
\rangle \nonumber \\ &=& \frac{1}{4} \langle
(-\underline{E}^{\dagger} P_R \tau_2+
\underline{E^{\prime}}^{\dagger} P_R) P(\theta_3)
(-\tau_2 P_R \underline{E} + P_R
\underline{E}^{\prime}) \nonumber \\ &&
(\underline{E}^{\dagger} P_R -
\underline{E^{\prime}}^{\dagger} P_R \tau_2)
P(\theta_4) (P_R \underline{E} - \tau_2 P_R
\underline{E}^{\prime}) \rangle. 
\label{gamma22_mz}
\end{eqnarray} Let
us assume the same statistical properties for
$\underline{E},\underline{E}^{\prime}$ here as for
$\underline{E}^{(1)},\underline{E}^{(2)}$ in
Section~\ref{hbt}: independent, time stationary, centred
random phase unpolarized 
%% page 14
Gaussian ensembles. Then
Eqns.(\ref{ensemble_properties1},\ref{ensemble_properties2}) are again
valid with
$\underline{E}^{(1)},\underline{E}^{(2)}
\longrightarrow
\underline{E},\underline{E}^{\prime}$. In
expression~(\ref{gamma22_mz}) there are again sixteen terms of
which (as in Section~\ref{hbt}) only 1111, 1122, 1221,
2112, 2211 and 2222 are nonzero. These have the
values: \begin{eqnarray}
1111&=&\frac{\kappa^2}{8}; \quad 1122=\frac{\kappa
\kappa^{\prime}}{16};  \nonumber \\
1221&=&\frac{\kappa \kappa^{\prime}}{4} {\rm Tr}
\left(P_R
P(\pi/2-\theta_3) P_L P(\theta_4)\right);
\nonumber \\ 2112 &=& (1221)^{\star}; \nonumber \\
2211&=&\frac{\kappa \kappa^{\prime}}{16}; \quad
2222=\frac{{\kappa^{\prime}}^2}{8}. \label{15}\end{eqnarray}
The final result for the HBT correlations is:
\begin{eqnarray} 
&&{\rm Tr}(P_R P(\pi/2-\theta_3)
P_L P(\theta_4)) = -\frac{1}{4} e^{2 i
(\theta_3+\theta_4)}, \nonumber \\ 
&&
\Gamma^{(2,2)}
= \frac{1}{8} (\kappa^2 +
\kappa\kappa^{\prime}+{\kappa^{\prime}}^2) -
\frac{\kappa \kappa^{\prime}}{8}
\cos{2(\theta_3+\theta_4)}. \nonumber\\
\label{gamma22_mz_geom}
\end{eqnarray} The
similarity to the expressions and results in
Section~\ref{hbt} is evident. However since now the two
propagation directions are mutually perpendicular,
there is no privileged Poincar\'{e} sphere in the
problem. It is true that the phase of the trace in
Eq.~(\ref{gamma22_mz_geom}) is most simply viewed as the solid angle
of a lune on a two-sphere, but this
two-sphere is useful for calculational purposes
alone and cannot be identified in any compelling
manner with any physically meaningful Poincar\'{e}
sphere in the present setup. For this reason it
seems not possible to interpret the phase in
Eq.~(\ref{gamma22_mz_geom}) as a nonlocal Pancharatnam phase, indeed
as a geometrical phase at all. Added to this is
the fact that once again even a small change in
the statistical properties of $\underline{E}$ and
$\underline{E}^{\prime}$ is likely to alter the
result~(\ref{gamma22_mz_geom}) completely in structure.
%%page 15
\section{Concluding remarks} 
\label{conc}
The attempt in~\cite{sam_nl} to
bring together the ideas of HBT correlations and
Pancharatnam - or more generally geometric
- phases is a very appealing one, even though our
  analysis suggests that this has not been
achieved. In discussions of quantum measurement theory, 
the appearance of nonlocality is in connection with 
composite systems in entangled states of spatially 
separated subsystems and observations on them. Since 
in our classical treatment we obtain the same expression 
for HBT correlations as found in~\cite{sam_nl}, it is clear 
that there is no quantum nonlocality involved. Further the 
way in which the Pancharatnam-like solid angle
appears here does 
not depend on transporting a pure state of any system over 
a closed path in any parameter or state space; hence while 
it reveals an interesting feature of HBT correlations which 
have indeed been experimentally verified~\cite{Martin}, its 
interpretation as a Pancharatnam phase seems open
to question.

In our view, 
from a wider perspective, the situation under discussion  has
similarities to the van Cittert-Zernike theorem in
partial coherence theory
- even with a spatially incoherent source,
  propagation can produce partial coherence at the
amplitude level. In a similar manner, in the case
of HBT correlations too we see that even if the
source intensities are uncorrelated, after
propagation nontrivial HBT correlations can
develop, since each detector (as in Figures 1 and
2) receives inputs from each source. An element of
nonlocality here is quite natural since wave
propagation involves spreading in space. If
polarizers are placed on the paths of propagating
beams, the fact that they can influence the HBT
correlations is also quite natural. However for
this to be interpretable in geometric phase terms
it seems necessary to have an underlying complex
Hilbert space structure and the pure state
evolution concept in a physically significant
manner.  This feature is absent in the
experimental setups and theoretical analysis
described in Sections~\ref{hbt} and~\ref{mz}.  

While we have shown that an entirely classical
treatment suffices  to obtain the results
of~\cite{sam_nl}  there are states of radiation
which are genuinely non classical and for which
HBT correlations take values beyond the classical
range of possibilities. Such states can certainly
be used in the experimental setups of Sections II
and III.  A proper quantum mechanical treatment
capable of handling such states must respect the
fundamental commutation relations describing
photons. This analysis will be presented
elsewhere. 

%%%%%%%%%%%%%%%%%%%%%%%%%%%%%%%%%%% 
\appendix
\section{}
\label{appendix}
We collect here some elementary formulae
concerning $2\times 2$ polarization matrices
needed
in the text. The matrices $P_R,P_L,P(\theta)$ are
Hermitian projections onto corresponding
two-component column vectors:
\begin{eqnarray}
&&\vert R \rangle = \frac{1}{\sqrt{2}} \left(
\begin{array}{c} 1 \\ i \end{array}\right), 
\vert L \rangle = \frac{1}{\sqrt{2}} \left(
\begin{array}{c} 1 \\ -i \end{array}\right),
\vert \theta \rangle = \left(
\begin{array}{c} \cos \theta \\ \sin \theta
\end{array} \right);\nonumber \\
&&\langle R,L \vert \theta \rangle =
\frac{1}{\sqrt{2}} e^{\mp i \theta};\nonumber \\
&&P_R=\vert R \rangle \langle R \vert,\quad 
P_L=\vert L \rangle \langle L\vert,  \quad
P(\theta)=P(\theta + \pi)=\vert \theta \rangle
\langle \theta \vert. \nonumber \\
\end{eqnarray}
The $\underline{\tau}$ matrices of polarization
optics are a cyclic rearrangement of the Pauli
matrices $\underline{\sigma}$ of quantum mechanics:
\begin{equation}
\tau_1=\sigma_3, \quad 
\tau_2=\sigma_1, \quad 
\tau_3=\sigma_2.
\label{pol_pauli}
\end{equation}
In terms of them we have
\begin{eqnarray}
&&
P_R=\frac{1}{2}(1+\tau_3),\quad
P_L=\frac{1}{2}(1-\tau_3),
\nonumber \\
&&P(\theta) = 
\frac{1}{2}(1+\tau_1 \cos 2\theta + \tau_2 \sin
2\theta).
\label{polarizer_rep}
\end{eqnarray}
Thus the circular polarization projectors $P_R$,
$P_L$ correspond to the N and S poles, $(0,0, \pm
1)$, on the Poincar\'{e} sphere; while the linear
polarization projector $P(\theta)$ maps to $(\cos
2 \theta, \sin 2 \theta ,0)$ on  the equator.

The elementary traces needed in
Eqn.(\ref{gamma_in_terms}) are 
\begin{equation}
{\rm Tr} (P_R P(\theta)) =
{\rm Tr} (P_L P(\theta)) = 
\vert \langle R,L \vert \theta  \rangle
\vert^2=\frac{1}{2}.
\end{equation}
The nontrivial trace in Eqn.(\ref{gamma_in_terms}) is a complex
quantity:
\begin{eqnarray}
{\rm Tr}(P_R P(\theta_3) P_L
P(\theta_4))&=&\frac{1}{4} \exp{(i {\rm ~arg}
\Delta_4(\vert R\rangle, \vert
\theta_3\rangle,\vert L\rangle,\vert
\theta_4\rangle))},  \nonumber \\
\Delta_4(\vert R\rangle, \vert
\theta_3\rangle,\vert L\rangle,\vert
\theta_4\rangle)&=&
\langle R\vert \theta_3\rangle
\langle \theta_3\vert L \rangle
\langle L\vert \theta_4\rangle
\langle \theta_4 \vert R\rangle.
\end{eqnarray}
This $\Delta_4$ is a four-vertex Bargmann
invariant, whose phase is known to be (the
negative of) a geometric phase. More precisely,
${\rm ~arg}
\Delta_4(\vert R\rangle, \vert
\theta_3\rangle,\vert L\rangle,\vert
\theta_4\rangle)=-\frac{1}{2}\Omega$, where
$\Omega$ is the solid angle on the Poincar\'{e} sphere
enclosed by the meridian from N to S at polar angle
$2\theta_3$ followed by the meridian from $S$ to $N$
at polar angle $2\theta_4$ :
\begin{eqnarray}
&{\rm Tr} (P_R P(\theta_3) P_L P(\theta_4)) =
\frac{1}{4} e^{-i \Omega/2},& \nonumber \\
&\Omega = 4 (\theta_3-\theta_4).&
\end{eqnarray}
In the scheme of Fig.~\ref{mz_setup} in
Section~{III}, reflection at a mirror amounts to
action on two-component electric field vectors
$\underline{E}$, $\underline{E}^{\prime}$ by the
matrix $-\tau_2$. With respect to the polarization
matrices the properties of $\tau_2$ are:
\begin{equation}
\tau_2 P_R\tau_2=P_L, \tau_2 P_L\tau_2=P_R,
\tau_2 P(\theta) \tau_2 = P(\pi/2-\theta).
\end{equation}
These are relevant in connection with
Eqns.~(\ref{15},\ref{gamma22_mz_geom}).
\begin{acknowledgments} 
Arvind acknowledges
funding from DST India under Grant No.
EMR/2014/000297.  NM thanks the Indian National
Science Academy for enabling this work through the
INSA C V Raman Research Professorship.
\end{acknowledgments}
%%%%%%%%%%%%%%%%%%%%%%%%%%%%%%%%%%%%%%%%%%%%%%%%%

%\bibliography{nl_geom}
\end{document}